\documentstyle[preprint,aps,floats]{revtex}
\begin{document}
\def\mh{m_h^{}}
\def\gev{\rm GeV}
\def\fbi{\rm fb^{-1}}
\def\ww{W^*W^*}
\def\zz{Z^*Z^*}
\def\lsim{\mathrel{\raise.3ex\hbox{$<$\kern-.75em\lower1ex\hbox{$\sim$}}}}
\def\gsim{\mathrel{\raise.3ex\hbox{$>$\kern-.75em\lower1ex\hbox{$\sim$}}}}
\def\ljj{\ell\nu jj}
\def\ll{\ell\bar\nu \bar\ell \nu}
\def\lljj{\ell\bar\ell jj}
\def\jj{\protect jj}
\def\llnn{\ell\bar\ell\nu\bar\nu}

\newcommand{ \slashchar }[1]{\setbox0=\hbox{$#1$}   
   \dimen0=\wd0                                     
   \setbox1=\hbox{/} \dimen1=\wd1                   
   \ifdim\dimen0>\dimen1                            
      \rlap{\hbox to \dimen0{\hfil/\hfil}}          
      #1                                            
   \else                                            
      \rlap{\hbox to \dimen1{\hfil$#1$\hfil}}       
      /                                             
   \fi}                                             %

\def\ptmiss{\slashchar{p}_{T}}
\def\etmiss{\slashchar{E}_{T}}
\input{epsf}
\ifx\epsffile\undefined
\message{(Uncomment input epsf to include figures)}
\newlength{\epsfysize}
\def\epsffile#1#2#3#4]#5{}
\fi
\tighten
\preprint{ \vbox{
\hbox{MADPH--98--1094}
\hbox{hep-ph/9812275}}}
\draft
\title{Exploiting $h \to \ww$ Decays\\  at the Upgraded Fermilab Tevatron}
\author{Tao Han$^1$, Andr\'e S. Turcot$^2$ and Ren-Jie Zhang$^1$}
\address{$^1$Department of Physics, University of Wisconsin\\ 
1150 University Avenue, Madison, WI 53706, USA\\
$^2$Department of Physics, University of Michigan,
Ann Arbor, MI 48109, USA}
\date{December, 1998}

\maketitle

\begin{abstract}
We study the observability of a Standard Model-like Higgs boson
at an upgraded Fermilab Tevatron via the mode $h \to \ww$. 
We concentrate on the main channel $gg\to h \to\ww\to \ll$.
We also find the mode
$q\bar q'\to W^\pm h \to W^\pm \ww\to \ell^\pm\nu\ell^\pm\nu jj$ useful.
We perform detector level simulations 
by making use of a Monte Carlo program SHW. 
Optimized searching strategy and kinematical cuts are developed. 
We find that with a c.~m.~energy of 2 TeV and an integrated
luminosity of 30 fb$^{-1}$ the signal should be observable at a 
3$\sigma$ level or better for the mass range of 
$145\ {\gev} \lsim \mh \lsim 180$ GeV. For 95\% 
confidence level exclusion, the mass reach is
$135\ {\gev} \lsim \mh \lsim 190$ GeV. 
We also present results of studying these channels with
a model-independent parameterization. Further improvement 
is possible by including other channels. We conclude that 
the upgraded Fermilab Tevatron will have the potential to 
significantly advance our knowledge of Higgs boson physics.
\end{abstract}

\section{Introduction}

The mass generation mechanisms for electroweak gauge
bosons and for fermions are among the most prominent
mysteries in contemporary high energy physics. 
In the Standard Model (SM) and its supersymmetric 
(SUSY) extensions, elementary scalar doublets of the
SU$_L$(2) interactions are responsible for the mass
generation. The scalar Higgs bosons are thus crucial 
ingredients in the theory, and searching for the Higgs 
bosons has been one of the major motivations in the 
current and future collider programs \cite{review}.
Although the masses of Higgs bosons are free parameters 
in the models, they are subject to generic bounds
based on theoretical arguments. The triviality bound
indicates that the Higgs boson mass ($m_h$) 
should be less than about 800 GeV for the SM to be 
a consistent low-energy effective theory \cite{triviality}. 
Vacuum stability argument, on the other hand, suggests a 
correlation between 
the $m_h$ lower bound and the new physics scale $\Lambda$ 
beyond which the SM is no longer valid \cite{vacuum}. 
In other words, the discovery
of SM-like Higgs boson implies a scale $\Lambda$ at
which new physics beyond the SM must set in;
and the smaller $\mh$ is, the lower $\Lambda$ would be. 
In the minimal supersymmetric Standard Model (MSSM), 
it has been shown that the mass of the lightest neutral Higgs boson 
must be less than about 130 GeV \cite{hmass},
and in any weakly coupled SUSY theory 
$m_h$ should be lighter than about 150 GeV \cite{mhbound}.
On the experimental side, the non-observation of Higgs signal
at the LEP2 experiments has established a lower bound
on the SM Higgs boson mass of 89.8 GeV at a 95\%
Confidence Level (CL) \cite{lep2b}. 
Future searches at LEP2 will eventually be
able to discover a SM Higgs boson with a mass up to
105 GeV \cite{lep2}. The CERN Large Hadron Collider
(LHC) is believed to be able to cover up to the full $m_h$
range of theoretical interest, about 1000 GeV \cite{lhc},
although it will be challenging to discover a Higgs boson
in the ``intermediate'' 
mass region 110 GeV $<m_h<$ 150 GeV, due to the huge SM
background to $h\to b\bar b$ and the requirement
of excellent di-photon mass resolution 
for the $h\to \gamma\gamma$ signal.

More recently, it has been discussed intensively how much the 
Fermilab Tevatron upgrade can do for the Higgs boson 
search \cite{run2}.
It appears that the most promising processes continuously 
going beyond the LEP2 reach would be the associated production 
of an electroweak gauge boson and the Higgs boson \cite{scott,GH,steve}
\begin{equation}
 p\bar p \to W hX,\ ZhX.
\label{whzh}
\end{equation}
The leptonic decays of $W,Z$ provide a good trigger and 
$h\to b\bar b$ may be reconstructible with adequate 
$b$-tagging. It is now generally believed that
for an upgraded Tevatron with a c.~m.~energy $\sqrt s=2$ TeV 
and an integrated luminosity ${\cal O}(10-30)\ \fbi$
a SM-like Higgs boson can be observed at a $3-5\sigma$
level up to a mass of about 120 GeV \cite{snowmass}.
The Higgs discovery through these channels crucially depends 
up on the $b$-tagging efficiency and the $b\bar b$ mass resolution. 
It is also limited by the event rate for $\mh > 120$ GeV. 
It may be possible to extend the mass reach to about
130 GeV by combining leptonic $W,Z$ decays
\cite{run2} and slightly beyond 
via the decay mode $h\to \tau^+\tau^-$ \cite{steve}. 
It is interesting to note that this mass reach is just near
the theoretical upper bound in MSSM. In the context of a 
general weakly-coupled SUSY model, it would be of great
theoretical significance for the upgraded Tevatron to extend 
the Higgs boson coverage to $\mh\sim 150$ GeV. Moreover,
it would have interesting implications on our knowledge for
a new physics scale $\Lambda$ if we do find a SM-like Higgs 
boson or exclude its
existence in the mass range 130 GeV$-$180 GeV, the
so-called ``chimney region'' between the triviality upper
bound and the vacuum stability lower bound \cite{upperb}.

\begin{figure}[tb]
\epsfysize=3.0in
\vskip 0.1in
\epsffile[5 220 300 515]{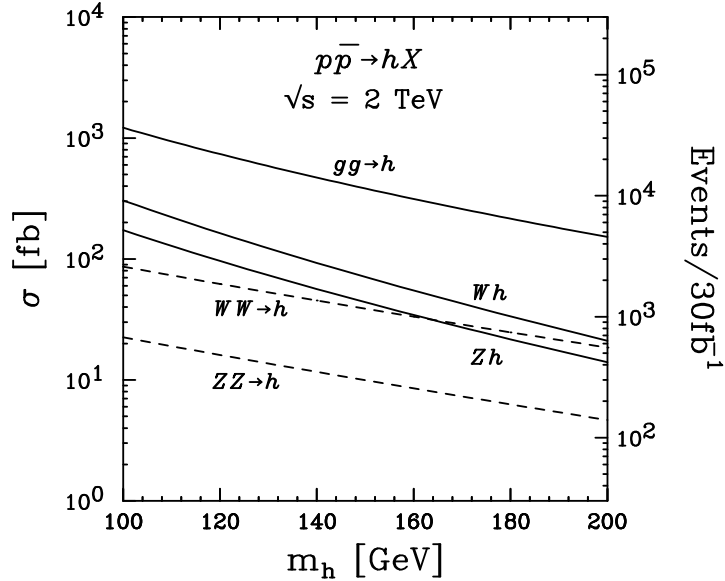}
\caption[]{The leading Higgs boson production cross sections
(in fb) versus $m_h$ at the $2$ TeV Tevatron. The solid curves are
for $gg \to h,\ q\bar q' \to W^\pm h$ and $q\bar q \to Z h$.
The dashed curves are for $W^+W^-$ and $ZZ$ fusion to $h$.
The scale on the right-hand side indicates the number of
events per 30 fb$^{-1}$ integrated luminosity.
QCD corrections \cite{spira,hw,hvw} have been included.
\label{one}}
\end{figure}

It is important to note that the leading production mechanism 
for a SM-like Higgs boson at the Tevatron is the gluon-fusion
process via heavy quark triangle loops
\begin{equation}
p \bar p \to gg X\to h X.
\label{gluon}
\end{equation}
There are also contributions to $h$ production from the vector
boson fusion processes\footnote{Here and henceforth, $W^*(Z^*)$ 
generically denotes a $W(Z)$ boson of either on- or off-mass-shell.}
\begin{equation}
\ww,\ \zz \to h,
\label{wwzz}
\end{equation}
where $\ww$ and $\zz$ are radiated off the quark partons.
In Fig.~\ref{one}, we present cross sections for SM Higgs
boson production at the Tevatron for processes (\ref{whzh}),
(\ref{gluon}) and (\ref{wwzz}). We see that the gluon fusion
process yields the largest cross section, typically a factor
of four above the associated production (\ref{whzh}). For $\mh>160$
GeV, $WW,ZZ$ fusion processes become comparable to that of 
(\ref{whzh}). In calculating the total cross sections, 
the QCD corrections have been 
included for all the processes \cite{spira,hw,hvw}, and
we have used the CTEQ4M parton distribution functions \cite{cteq4m}.

\begin{figure}[tb]
\epsfysize=3.6in
\epsffile[37  230  560  515]{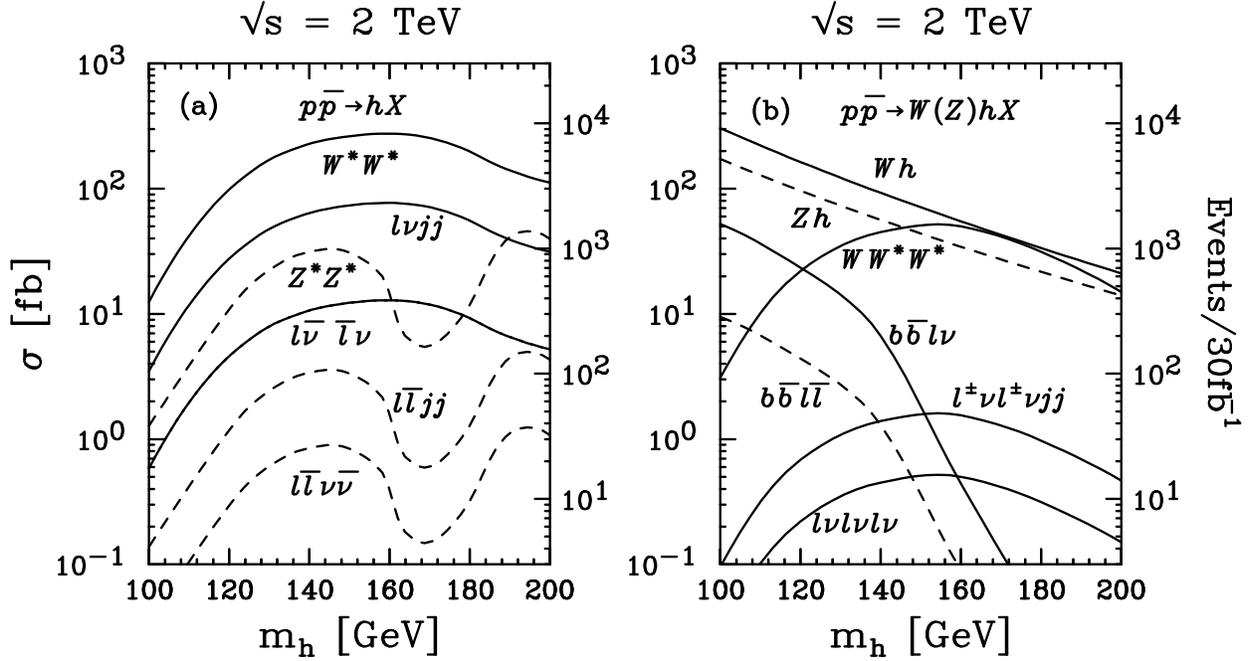}
\vskip 0.2in
\caption[]{The Higgs boson production cross sections
(in fb) and various subsequent decay modes
versus $m_h$ at the $2$ TeV Tevatron
for (a) $gg \to h \to \ww$ (solid) 
and $\zz$ (dashed), (b) $q\bar q' \to Wh$ with $h \to \ww$ (solid) 
and $Zh$ (dashed). Also shown are $h \to b\bar b$ 
with $W,Z$ leptonic decays.
The scales on the right-hand side indicate the number of
events per 30 fb$^{-1}$ integrated luminosity.
\label{two}}
\end{figure}

Although the decay mode $h\to b\bar b$ in Eqs.~(\ref{gluon}) 
and (\ref{wwzz}) would be swamped by the QCD background, 
the decay modes to vector boson pairs 
\begin{equation}
 h\to W^*W^*,\ Z^*Z^*
\label{vv}
\end{equation}
will have increasingly large branching fractions
for $m_h\gsim 130$ GeV and are natural channels to consider
for a heavier Higgs boson. 
In Fig.~\ref{two}(a), we show the cross sections for $gg\to h$
with $h\to W^*W^*$ and $Z^*Z^*$ versus $\mh$ at $\sqrt s=2$ TeV. 
The leptonic decay channels are also separately shown 
by solid and dashed curves, respectively, for
\begin{eqnarray}
 h \to && \ww \to \ell\nu jj\ \ {\rm and}\ \ 
\ell\bar\nu \bar\ell \nu ,\label{sigww}\\ 
       && \zz \to \ell\bar \ell jj\ \ {\rm and}\ \ 
\ell\bar \ell \nu\bar \nu ,
\label{sigzz}
\end{eqnarray}
where $\ell=e,\mu$ and $j$ is a quark jet.
The scale on the right-hand side gives the number of events 
expected for 30 $\fbi$. We see that for the $\mh$ range of current
interest, there will be about 1000 events produced for 
the semi-leptonic mode $\ww \to \ljj$ and about 300 events 
for the pure leptonic mode $\ww \to \ll$.
Although the $\ljj$ mode has a larger production rate, the
$\ll$ mode is cleaner in terms of the SM background contamination.
The corresponding modes from $\zz$ are smaller by about an 
order of magnitude.
It is natural to also consider the $h\to \ww$ mode from the 
$Wh$ associated production in Eq.~(\ref{whzh}).
This is shown in Fig.~\ref{two}(b) by the solid curves for 
\begin{eqnarray}
W^\pm h \to \ell^\pm \nu\  \ww \to 
&& \ell \nu\ \ell\nu\  \ell\nu,
\label{unlikesign}\\ 
&& \ell^\pm \nu\ \ell^\pm \nu\  jj.
\label{likesign}
\end{eqnarray}
The trilepton signal is smaller than the like-sign lepton plus jets signal
by about a factor of three due to the difference of $W$ decay branching 
fractions to $\ell=e,\mu$ and to jets. 
For comparison, also shown in Fig.~\ref{two}(b) 
are $Wh\to b\bar b\ell\nu $ (solid) 
and $Zh\to  b\bar b\ell\bar \ell$ (dashed) via $h\to b\bar b$. 
We see that the signal rates for these
channels drop dramatically for a higher $\mh$. 
Comparing the $h$ decays in Fig.~\ref{two}(a) and (b), 
it makes the gauge boson pair modes of Eq.~(\ref{vv}) a clear choice 
for Higgs boson searches beyond 130 GeV.

In fact, the pure leptonic channel in Eq.~(\ref{sigww})
has been studied at the SSC and LHC energies \cite{lnulnu,herbi} 
and at a 4 TeV Tevatron \cite{GH}. Despite the difficulty in
reconstructing $\mh$ from this mode due to the two missing
neutrinos, the obtained results for the signal identification
over the substantial SM backgrounds were all encouraging. 
In a more recent paper \cite{hz}, 
two of the current authors carried out a parton-level
study for the $\ww$ channels of Eq.~(\ref{sigww}) for the
2 TeV Tevatron upgrade. We found that the di-lepton mode
in Eq.~(\ref{sigww}) is more promising than that of
$\ell\nu jj$ due to the much larger QCD background
to the latter. While the results were encouraging, realistic
simulations including detector effects were
called for to draw further conclusions. 

In this paper, we concentrate on the pure leptonic channel
and carry out more comprehensive analyses for the signal
and their SM backgrounds. 
We perform detector level simulations by making use of
a Monte Carlo program SHW developed for the Run-II SUSY/Higgs
Workshop \cite{shw,run2}. 
We present optimized kinematic cuts which can adequately 
suppress the large SM backgrounds and, moreover, have been
structured so as to provide a statistically robust background
normalization.
For the $Wh\to W\ww$ channel, although the trilepton signal of 
Eq.~(\ref{unlikesign}) is rather weak, 
the like-sign leptons plus two jets in Eq.~(\ref{likesign})
can be useful to enhance the signal observability.
For completeness, we have also included the contributions 
from the vector boson fusion of Eq.~(\ref{wwzz}) and 
$W\to \tau\nu \to \nu \ell\nu_\ell$
decay mode, although they are small. 
We also comment on the systematic
effects on the signal and background measurements
which would degrade signal observability. 
After combining all the channels studied, we find that with a 
c.~m.~energy of 2 TeV and an integrated luminosity of 30 fb$^{-1}$,
the signal of $h\to \ww$ can be observable at a 
3$\sigma$ level or better for the mass range of 
$145\ {\gev} \lsim \mh \lsim 180$ GeV. For 95\% 
CL exclusion, the mass reach is
$135\ {\gev} \lsim \mh \lsim 190$ GeV. We thus conclude that
the upgraded Fermilab Tevatron will have the potential to 
significantly advance our knowledge of Higgs boson physics.
This provides strong motivation for luminosity 
upgrade of the Fermilab Tevatron beyond the Main Injector plan.

Our signal and background Monte Carlo simulation was performed
using the PYTHIA package \cite{pythia} interfaced with the
SHW detector simulation \cite{shw}. For pair production of
resonances, {\em e.g.} $WW$, PYTHIA incorporates
the full $2\rightarrow 2 \rightarrow 4$ matrix elements thereby
insuring proper treatment of the final state angular correlations.
Similarly for $h\to WW$, the angular correlations between the four final
state fermions have been taken into account. The full $Z/\gamma^*$ 
interference is simulated for $ZZ$ production; however, the $WZ$ 
process considers only the pure $Z$ contribution. 
For Higgs boson production in association with a
gauge boson in Eq~(\ref{whzh}), the associated 
$W$ and $Z$ decay angular distributions are treated properly.
The production cross-sections for the principal background processes 
were normalized to 
$\sigma(WW) = 10.4$ pb,  $\sigma(t\bar{t}) = 6.5$ pb, 
$\sigma(WZ) = 3.1$ pb, and $\sigma(ZZ) = 1.4$ pb.

The rest of the paper is organized as follows.
In sections II and III, we present in details our
studies for the pure leptonic and like-sign leptons plus jets
signals, respectively. In section IV, we first summarize our
results. We then present a study of these channels with
a model-independent parameterization for the signal cross section. 
We conclude with a few remarks.

\section{Di-leptons plus Missing Transverse Energy Signal}

\begin{figure}[tb]
\epsfysize=5.5in
\epsffile[-50 0 560 560]{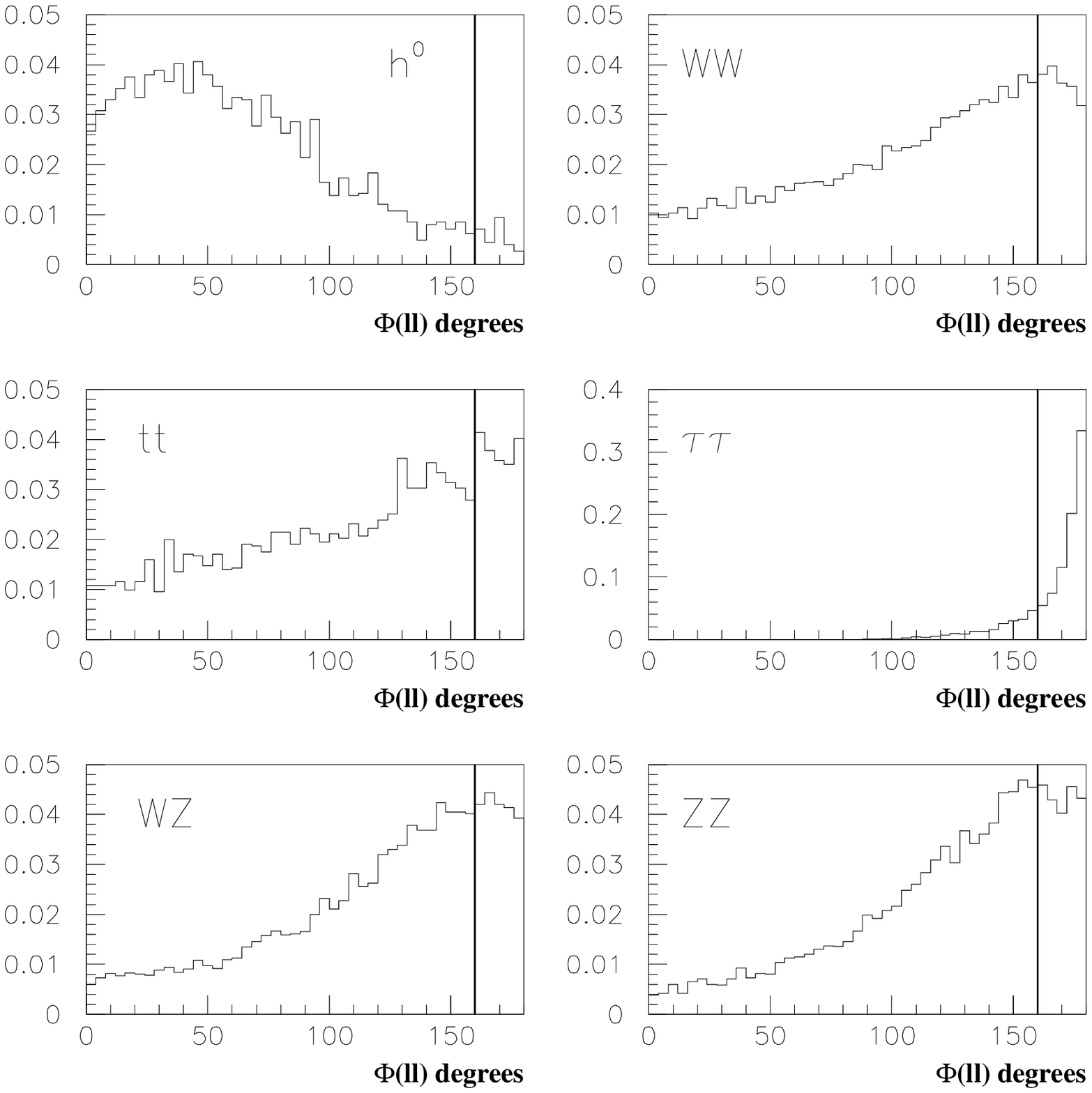} 
\caption[]{Normalized azimuthal angle distributions 
${1\over \sigma} {d\sigma\over d\phi(\ell\ell)}$ for the
signal $gg \to h \to \ww\break \to \ll$ with $\mh=170$ GeV
and backgrounds $WW$, $t \bar t$, $\tau^+\tau^-$, 
$WZ$ and $ZZ$. 
\label{phis}}
\end{figure}

\begin{figure}[thb]
\epsfysize=5.5in
\epsffile[-50 0 567 567]{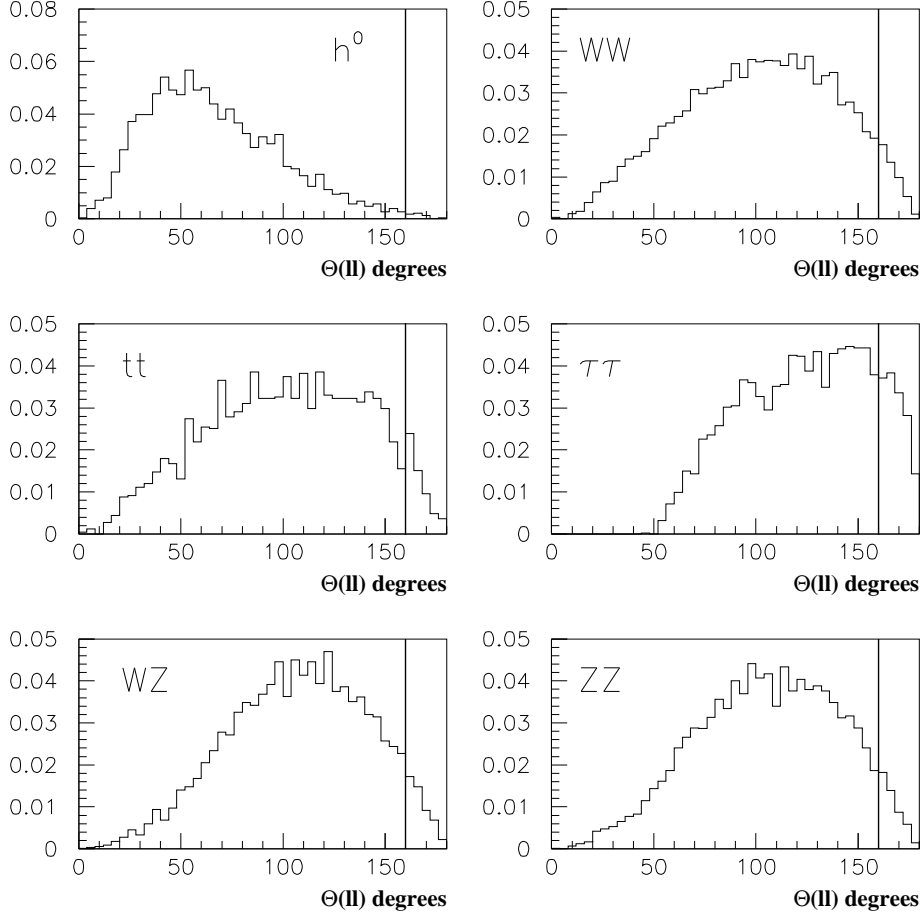} 
\caption[]{Normalized distributions 
${1\over \sigma} {d\sigma\over d\theta(\ell\ell)}$
for the opening angle in Eq.~(\ref{phi}) 
for the signal $gg \to h \to \ww \to \ll$ with $\mh=170$ 
GeV and backgrounds
$WW$, $t \bar t$, $\tau^+\tau^-$, $WZ$ and $ZZ$. 
\label{thes}}
\end{figure}

\begin{figure}[tb]
\epsfysize=5.5in
\epsffile[-50 0 560 560]{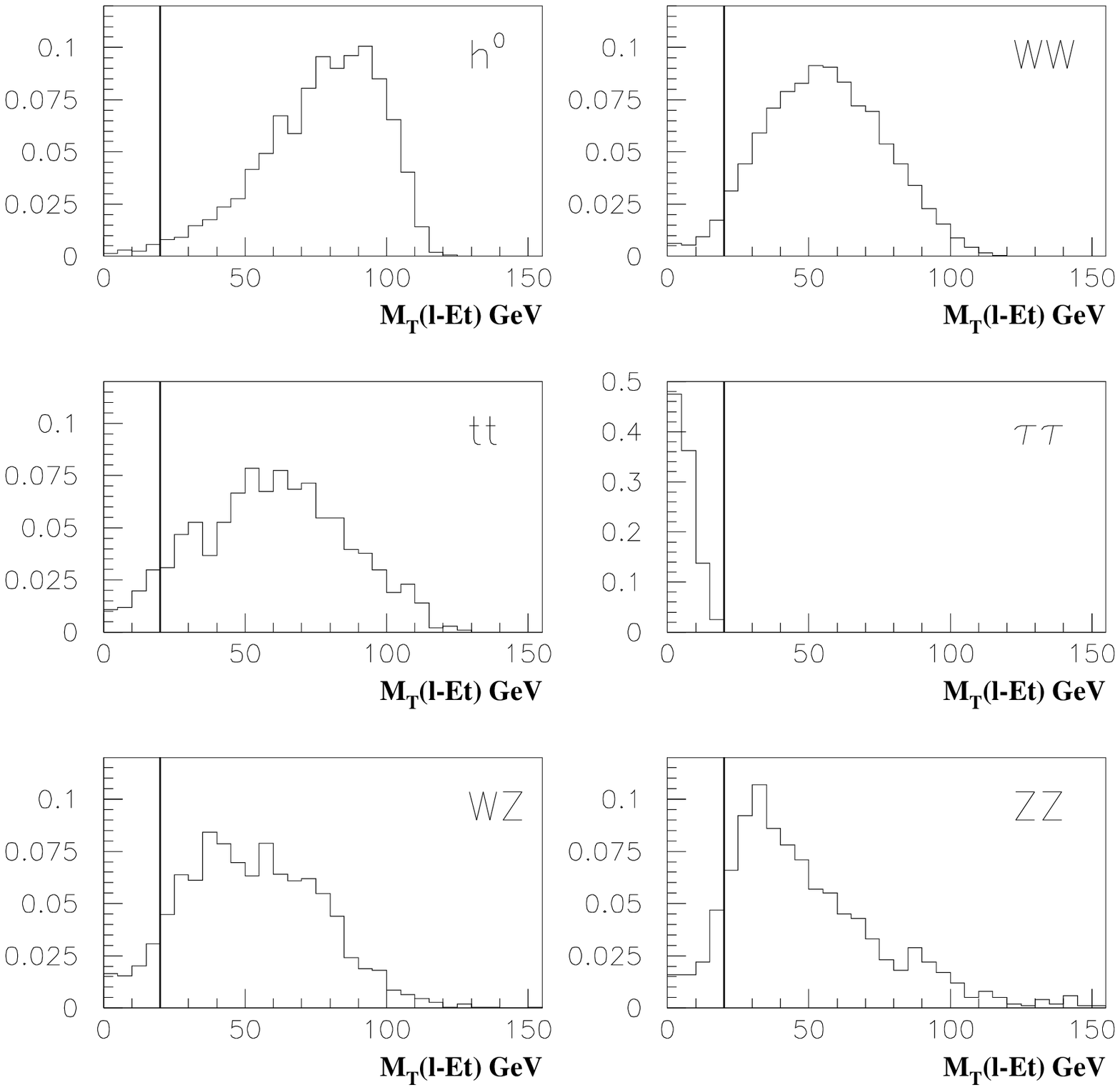}
\caption[]{Normalized distributions 
${1\over \sigma} {d\sigma\over dM_T}$ 
for the two-body transverse-mass defined in Eq.~(\ref{tm}) 
for the signal $gg \to h \to \ww \to \ll$ with $\mh=170$ GeV
and backgrounds $WW$, $t \bar t$, $\tau^+\tau^-$, $WZ$ and $ZZ$. 
The minimum of $M_T(\ell_1 \etmiss)$ and $M_T(\ell_2 \etmiss)$ is shown.
\label{mtlnu}}
\end{figure}

For the pure leptonic channel in Eq.~(\ref{sigww}), 
we identify the final state signal as two isolated 
opposite-sign charged leptons and 
large missing transverse energy. 
The leading SM background processes are
\begin{eqnarray}
p\bar p &\to& W^+W^-\to  \ell \bar \nu \bar \ell \nu,\ \ 
 Z Z(\gamma^*)\to \nu \bar \nu \ell \bar \ell,\ \  
 W Z(\gamma^*)\to \ell \bar \nu \ell \bar \ell,
\nonumber\\
p\bar p &\to& t \bar t\to  \ell \bar \nu \bar \ell \nu b\bar b,
\ \  
p\bar p \to Z(\gamma^*)\to  \tau^+\tau^- \to 
\ell \bar \nu \bar \ell \nu \nu_\tau \bar \nu_\tau .
\label{dy}
\end{eqnarray}
We first impose basic acceptance cuts for the 
leptons\footnote{The cuts for leptons were chosen to reflect
realistic trigger considerations. It is desirable to
extend the acceptance in $\eta_\ell^{}$.}
\begin{eqnarray}
\nonumber
&&p^{}_T(e)  > 10\ {\gev},\ \ |\eta^{}_e| < 1.5,\nonumber\\
&&p^{}_T(\mu_1)>10\ {\gev},\ \  p^{}_T(\mu_2)>5\ {\gev},\ \ 
|\eta^{}_\mu| < 1.5,\nonumber \\
&&m(\ell\ell)>10\ {\gev},\ \ \Delta R(\ell j)>0.4,\ \ 
\etmiss > 10\ {\gev},
\label{basic}
\end{eqnarray}
where $p^{}_T$ is the transverse momentum and $\eta$ the
pseudo-rapidity.
The cut on the invariant mass $m(\ell\ell)$ is to remove the
photon conversions and leptonic $J/\psi$ and $\Upsilon$ decays.
The isolation cut on $\Delta R(\ell j)$ removes the muon events 
from heavy quark ($c, b$) decays.\footnote{The electron identification
in the SHW simulation imposes strict isolation requirements already.} 

\begin{figure}[tb]
\epsfysize=5.5in
\epsffile[-50 0 567 567]{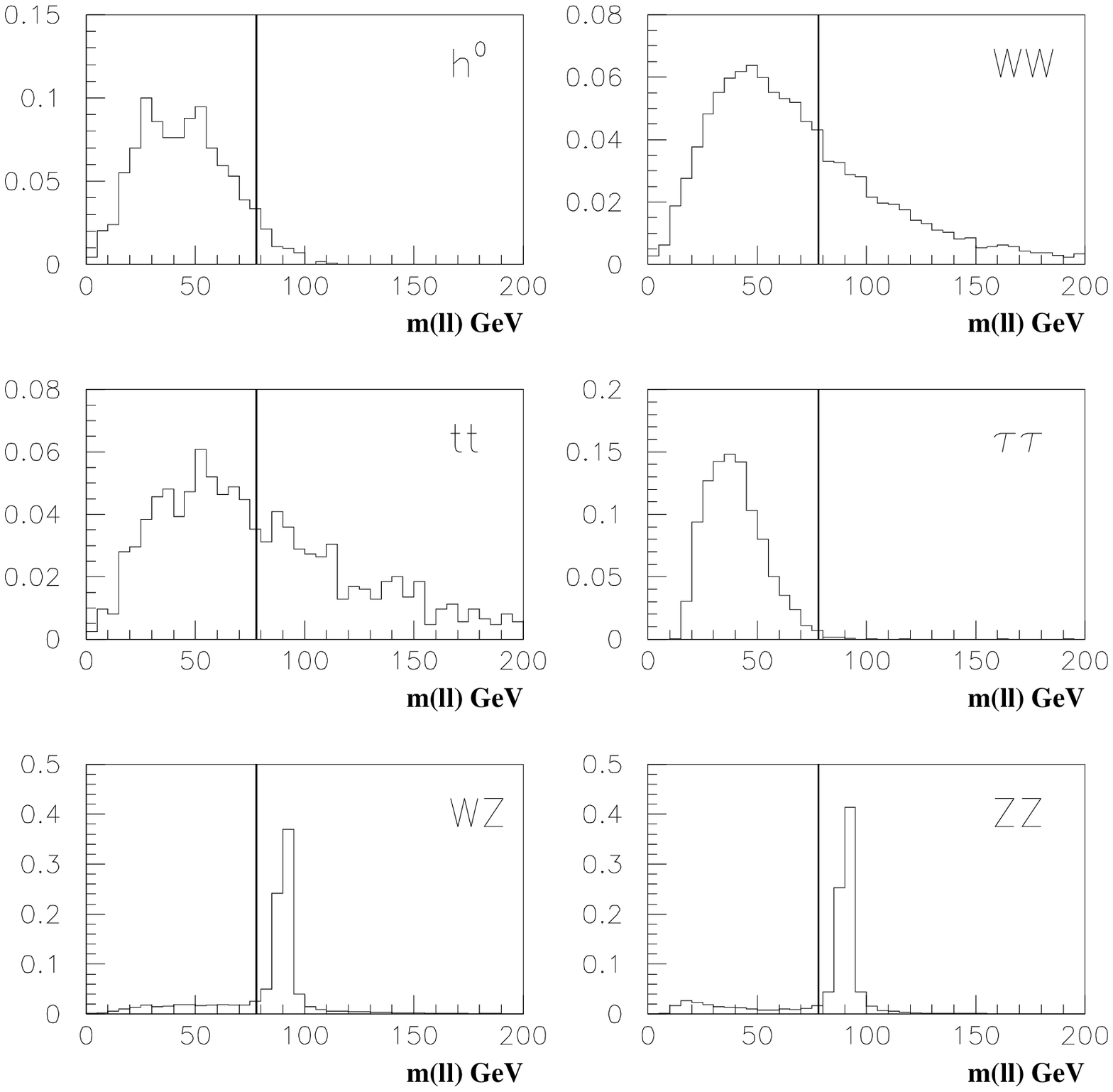}
\caption[]{Normalized like-flavor lepton-pair invariant mass distributions 
${1\over \sigma} {d\sigma\over dm(\ell\ell)}$ 
for the signal $gg \to h \to \ww \to \ll$ with $\mh=170$ GeV
and backgrounds $WW$, $t \bar t$, $\tau^+\tau^-$, $WZ$ and $ZZ$. 
\label{mlls}}
\end{figure}

At this level, the largest background comes from the Drell-Yan
process for $\tau^+\tau^-$ production. However, the charged
leptons in this background are very much back-to-back and this
feature is also true, although to a lesser extent, for other 
background processes as well.
On the other hand, due to the spin correlation of the Higgs boson
decay products, the two charged leptons tend to move in parallel.
We demonstrate this point in Figs.~\ref{phis} and \ref{thes}
where the distributions of the azimuthal angle in the transverse 
plane [$\phi(\ell\ell)$] and the three-dimensional opening-angle 
between the two leptons [$\theta(\ell\ell)$] for the signal 
and backgrounds are shown.\footnote{Since we are mainly interested
in the shapes of the kinematic distributions, we present them
normalized to unity with respect to the total cross section
with appropriate preceeding cuts.}
This comparison motivates us to impose the cuts 
\begin{equation}
\phi(\ell\ell) < 160^\circ,\ \ 
\theta(\ell\ell) < 160^\circ.
\label{phi}
\end{equation}
The $\tau^+\tau^-$ background can be essentially 
eliminated with the help of additional cuts
\begin{equation}
p_T^{}(\ell\ell) > 20\ {\gev},\ \ 
\cos\theta_{\ell\ell-\etmiss} < 0.5,\ \ 
M_T^{}(\ell \etmiss) >20\ {\gev},
\label{ptll}
\end{equation}
where 
$\theta_{\ell\ell-\etmiss}$ is the relative angle between the lepton 
pair transverse momentum and the missing transverse momentum, 
which is close to 180$^\circ$ for the signal and near 0$^\circ$
for the Drell-Yan $\tau^+\tau^-$ background. The
two-body transverse-mass is defined for each
lepton and the missing energy as 
\begin{equation}
M_T^2(\ell \etmiss)=2 p_T(\ell)\etmiss(1-\cos\theta_{\ell-\etmiss}),
\label{tm}
\end{equation}
and the distributions are shown in Fig.~\ref{mtlnu}.

\begin{figure}[tb]
\epsfysize=5.5in
\epsffile[-50 0 560 560]{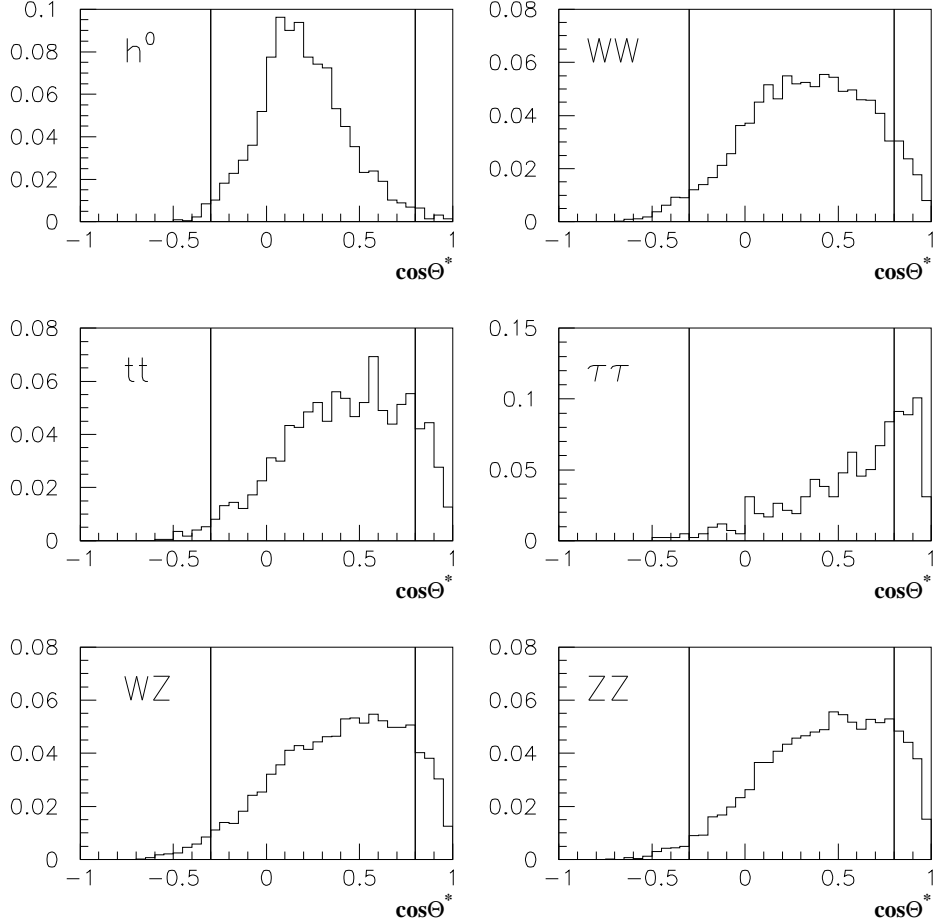}
\caption[]{Normalized distributions 
${1\over \sigma} {d\sigma\over d \cos\theta^*_{\ell_1} }$ 
for the correlation angle defined
above Eq.~(\ref{theta}) for the signal $gg \to h \to \ww \to \ll$ 
with $\mh=170$ GeV and backgrounds $WW$, $t \bar t$, $\tau^+\tau^-$, 
$WZ$ and $ZZ$. 
\label{thetas}}
\end{figure}

We can further purify the signal by removing the high
$m(\ell\ell)$ events from $Z\to \ell\bar \ell$ 
as well as from $t\bar t, W^+W^-$, 
as demonstrated in Fig.~\ref{mlls}. We therefore impose
\begin{eqnarray}
\nonumber
m(\ell\ell)&<& 78\ {\gev}\ \ {\rm for}\ e^+e^-,\mu^+\mu^-,\\ 
m(\ell\ell)&<& 110\ {\gev}\ \ {\rm for}\ e\mu.
\label{zout}
\end{eqnarray}
As suggested in Ref.~\cite{herbi}, the lepton correlation angle
between the momentum vector of the lepton pair and the momentum of
the higher $p_T$ lepton ($\ell_1$) in the lepton-pair rest frame, 
$\theta^*_{\ell_1}$, also has discriminating power between the
signal and backgrounds. This is shown in Fig.~\ref{thetas}. 
We thus select events with
\begin{equation}
-0.3 < \cos\theta^*_{\ell_1} < 0.8.
\label{theta}
\end{equation}
A characteristic feature of the top-quark background is the presence
of hard $b$-jets. We thus devise the following jet-veto 
criteria\footnote{The previous study \cite{hz} at the parton-level 
suggested a more stringent jet-veto cut. 
It turns out that it would be too costly
for the signal and the more sophisticated jet-veto criteria of
Eq.~(\ref{jetveto}) is thus desirable.}
\begin{eqnarray}
{\rm veto\ if}\ \ p_T^{j_1}>95\ {\gev},\ \  |\eta^{}_j| < 3,\nonumber\\ 
{\rm veto\ if}\ \ p_T^{j_2}>50\ {\gev},\ \  |\eta^{}_j| < 3,\nonumber\\ 
{\rm veto\ if}\ \ p_T^{j_3}>15\ {\gev},\ \  |\eta^{}_j| < 3.
\label{jetveto}
\end{eqnarray}
Furthermore, if either of the two hard jets ($j_1,j_2$) is identified
as a $b$ quark, the event will be also vetoed. The $b$-tagging 
efficiency is taken to be \cite{shw}
\begin{equation}
\epsilon_b = 1.1\times 57\% \  
{\rm tanh}({\eta_b\over 36.05}),
\label{btag}
\end{equation}
where the factor 1.1 reflects the 10\%
improvement for a lepton impact parameter tag.

The results up to this stage are summarized in Table~\ref{tabI} 
for the signal $\mh=140-$190 GeV as well as the SM backgrounds. 
The acceptance cuts discussed above are fairly efficient, 
approximately 35\% 
in retaining the signal while suppressing the backgrounds
substantially. We see that the dominant background comes
from the electroweak $WW$ production, about a factor of 30
higher than the signal rate. The sub-leading backgrounds $t\bar t$
and $W$+fake (the background where a jet mimics an electron 
with a probability of $P(j\to e)=10^{-4}$ \cite{fake})
are also bigger than the signal.
We note that although the $b$-jet veto is
effective against the $t\bar t$ background, the final results
are not affected if the veto efficiency is significantly
worse.

\begin{table}[t]
\begin{tabular}{|l|c|c|c|c|c|c|}
 $\mh$ [GeV] & 140 & 150 & 160 & 170 & 180   & 190\\  \hline
 signal [fb] & 3.9 & 4.4 & 5.2 & 4.8 & 3.6   & 2.5\\  \hline
 effic. [\%] & 35  & 34  & 38  & 39  & 36    & 37\\   \hline \hline
{}&$WW$ & $t\bar t$ & $\tau^+\tau^-$& $WZ$ & $ZZ$ & $W+$fake\\ 
\hline
bckgrnds [fb]& 130 & 13  & 0   & 4.4 & 2.4   & 18 \\ 
\end{tabular}
\vspace{0.2in}
\caption[]{$h\rightarrow \ww \rightarrow \ll$ signal cross section
(in fb) for $\mh=140-$190 GeV and various SM backgrounds after 
the kinematical cuts of Eqs.~(\ref{basic})$-$(\ref{jetveto}). 
The signal efficiencies are also shown (in percentage). 
$W+$fake refers to the background where a jet mimics
an electron with a probability of $P(j\to e)=10^{-4}$.
The backgrounds are independent of $\mh$.
}
\label{tabI}
\end{table}

\begin{figure}[tb]
\epsfysize=4.5in
\epsffile[-100 0 560 560]{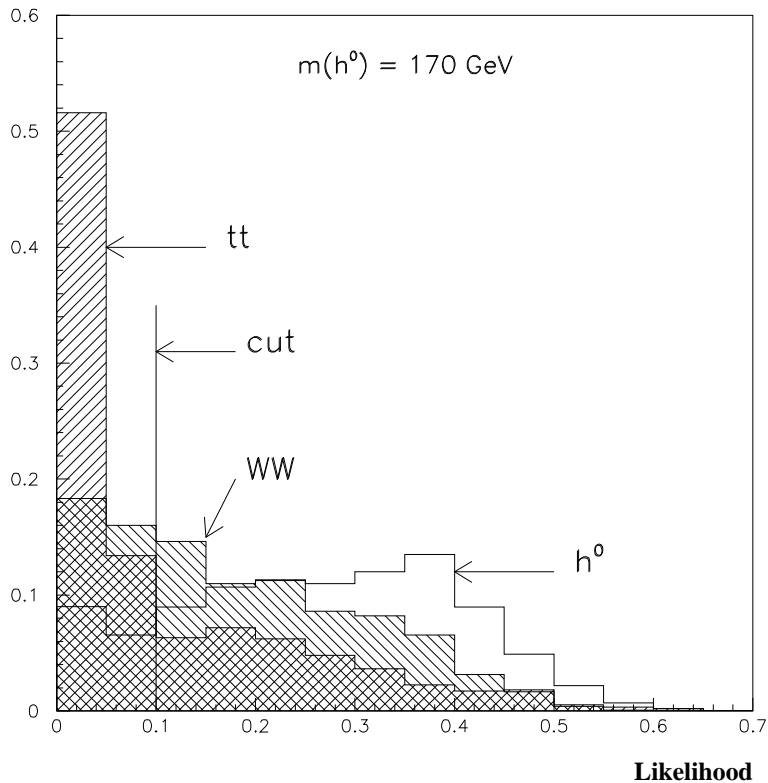}
\caption[]{Distributions for the likelihood variable
defined in Eq.~(\ref{like}) for the signal $\mh=170$ GeV
and the leading SM backgrounds $WW$ and $t\bar t$.
\label{likeli}}
\end{figure}

One can improve the signal observability by
constructing a likelihood based on some characteristic
kinematical variables. We choose the variables as
\begin{itemize}
\item[1.] $\cos\theta_{\ell\ell}$, the polar angle with respect 
to the beam axis of the di-lepton \cite{herbi};
\item[2.] $\phi(\ell\ell)$ as in Eq.~(\ref{phi});
\item[3.] $\theta(\ell\ell)$ as in Eq.~(\ref{phi});
\item[4.] $\cos\theta_{\ell\ell-\etmiss}$ as in Eq.~(\ref{ptll});
\item[5.] $p_T^{j1}$ as in Eq.~(\ref{jetveto});
\item[6.] $p_T^{j2}$ as in Eq.~(\ref{jetveto}).
\end{itemize}
We wish to evaluate the likelihood for a candidate event
to be consistent with one of five event classes: a Higgs
boson signal ($140 < \mh <$ 190 GeV), 
$WW$, $t\bar t$, $WZ$ or $ZZ$. For a single variable $x_i$, 
the probability for an event to belong to class $j$ is given by
\begin{equation}
P^j_i(x_i) = \frac{f^j_i(x_i)}{\Sigma_{k=1}^5\ f^k_i(x_i)},
\end{equation}
where $f_i^j$ denotes the probability density for 
class $j$ and variable $i$. 
The likelihood of an event to belong
to class $j$ is given by the normalized products of the individual
$P^j_i(x_i)$ for the $n=6$ kinematical variables:
\begin{equation}
{\cal L}^j = \frac{ \Pi_{i=1}^n\  P^j_i(x_i)}
{\Sigma_{k=1}^5\ \Pi_{i=1}^n\  P^k_i(x_i)},
\label{like}
\end{equation}
The value of ${\cal L}^j$ for a Higgs boson signal hypothesis 
($j=1$) is shown in Fig.~\ref{likeli} 
where it can be seen that a substantial fraction
of the $t\bar t$ and $WW$ background can be removed for a modest loss
of acceptance. The $WZ$ and $ZZ$ backgrounds
have similar distributions to the
$WW$ and have been omitted for clarity. 
We thus impose the requirement
\begin{equation}
{\cal L}^{j=1} > 0.10.
\label{likec}
\end{equation}
The improved results are summarized in Table~\ref{tabII}.

\begin{table}[ht]
\begin{tabular}{|l|c|c|c|c|c|c|}
 $\mh$ [GeV] & 140 & 150 & 160 & 170 & 180   & 190\\  \hline
 signal [fb] & 3.1 & 3.6 & 4.5 & 4.1 & 2.9   & 2.0\\  \hline
\hline
{}&$WW$ & $t\bar t$ & $\tau^+\tau^-$& $WZ$ & $ZZ$ & $W+$fake\\ 
\hline
bckgrnds [fb]& 83 & 4.5  & 0   & 3.1 & 1.8   & 13 \\ 
\end{tabular}
\vspace{0.2in}
\caption[]{$h\rightarrow \ww \rightarrow \ll$ signal cross section
(in fb) for $\mh=140-$190 GeV and various SM backgrounds after the kinematical 
cuts of Eqs.~(\ref{basic})$-$(\ref{jetveto}) and the likelihood cut
Eq.~(\ref{likec}). $W+$fake refers to the background where a jet mimics
an electron with a probability of $P(j\to e)=10^{-4}$.
The backgrounds are independent of $\mh$.
}
\label{tabII}
\end{table}

\begin{figure}[tb]
\epsfysize=5.5in
\epsffile[-20 80 567 567]{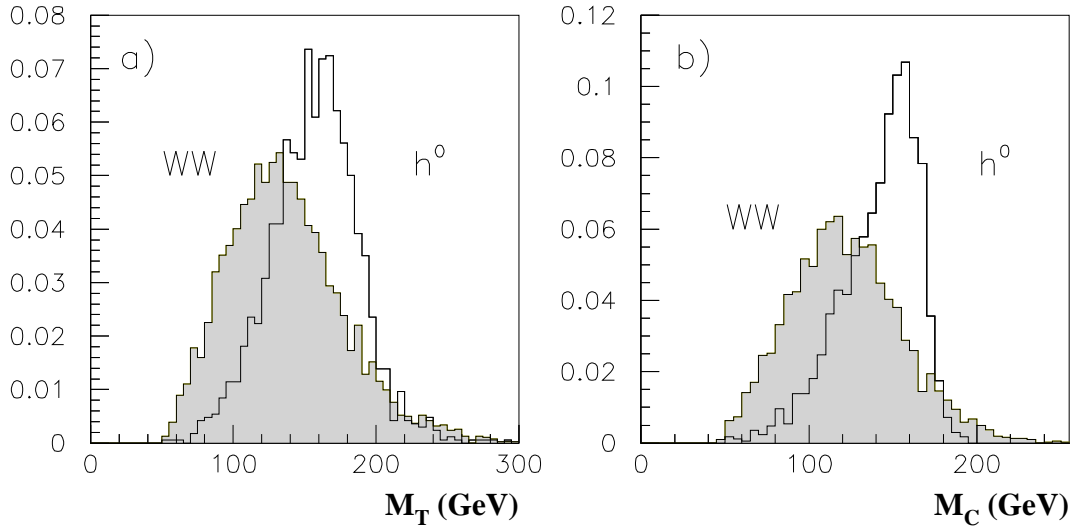}
\vskip -2.0in
\caption[]{Normalized distributions 
${1\over \sigma} {d\sigma\over d M}$ for the signal 
$gg \to h \to \ww \to \ll$ with $\mh=170$ GeV (histogram)
and the leading $WW$ background (shaded) for (a) the transverse mass
defined in Eq.~(\ref{mt}),
and (b) the cluster transverse mass defined in 
Eq.~(\ref{mc}).
\label{masses}}
\end{figure}

In identifying the signal events, it is crucial to
reconstruct the mass peak of $\mh$. Unfortunately,
the $\ww$ mass from the $h$ decay cannot be accurately 
reconstructed due to the two undetectable neutrinos. 
However, both the transverse mass $M_T$ and the 
cluster transverse mass $M_C$ \cite{bho}, defined as
\begin{eqnarray}
M_T &=& 2\sqrt{ p^2_T(\ell\ell)+m^2(\ell\ell)},\label{mt} \\ 
M_C &=&  \sqrt{ p^2_T(\ell\ell)+m^2(\ell\ell)}\ + \etmiss,
\label{mc}
\end{eqnarray}
yield a broad peak near $\mh$ and have a long tail below.
The cluster transverse mass $M_C$ has a Jacobian structure
with a well defined edge at $m_h$.
We show the nature of these two variables for the signal
with $\mh=170$ GeV and the leading $WW$ background
in Fig.~\ref{masses}(a) for $M_T$ and (b) for $M_C$ after application
of the likelihood cut.
For a given $m_h$ to be studied, one can perform additional cut optimization.
In Table~\ref{tabcut}, we list $m_h$-dependent criteria for
the signal region defined as
\begin{equation}
\mh - 60 < M_C < \mh + 5\ \gev.
\end{equation}
\begin{table}[b]
\begin{tabular}{|l|c|c|c|c|c|c|}
 $\mh$ [GeV] & 140 & 150 & 160 & 170 & 180   & 190\\  \hline
 $\cos\theta^*_{\ell_1}$ & - & $<$0.6 & 0.35 & 0.35 & 0.55   & 0.75\\ \hline
 $\etmiss$ & $>$25 & 25 & 30 & 35 & 40 & 40\\  \hline
min[$M_T^{}(\ell_1 \etmiss),M_T^{}(\ell_2 \etmiss)$]  
& $>$40 & 40 & 75 & 80 & 85   & 75\\  \hline
$M_T^{}(\ell_1 \etmiss)$ & $>$60 & 60 & - & - & - & -\\  \hline
$m(\ell\ell)$ & $<$65 & 65 & 65 & 75 & 85 & -\\  \hline
$p_T^{}(\ell\ell)$ & $>$40 & 50 & 65 & 70 & 70 & 70\\  \hline
$\theta(\ell\ell)$ & $<$100 & 100 & 70 & 70 & 90  & 90\\ \hline
$M_T^{}$ & - & $>$110 & 120 & 130 & 140 & 140\\
\end{tabular}
\vspace{0.2in}
\caption[]{Summary of the optimized cuts additional to those
in Eqs.~(\ref{basic})$-$(\ref{jetveto}) for various Higgs boson mass.
}
\label{tabcut}
\end{table}

\begin{figure}[thb]
\epsfysize=5.5in
\epsffile[-20 80 567 567]{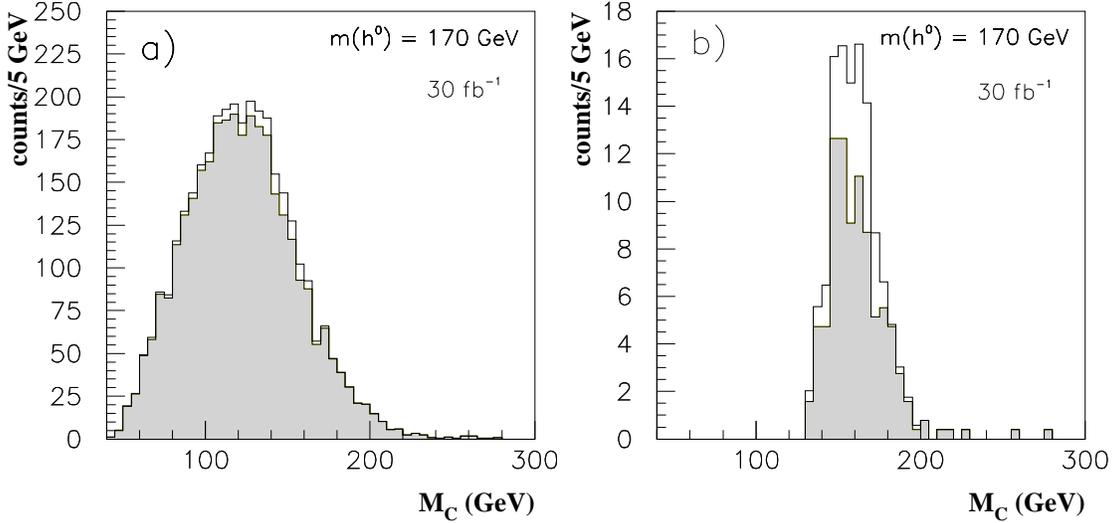}
\vskip -2.0in
\caption[]{Cluster transverse mass distributions for
the leading $WW$ background (shaded) and the background
plus the signal $gg \to h \to \ww \to \ll$ with $\mh=170$ GeV
(histogram) (a) before the optimized
cuts in Table~\ref{tabcut} and (b) after the cuts. The
vertical axis gives the number of events per 5 GeV bin
for 30 $\fbi$.
\label{bacuts}}
\end{figure}

We illustrate the effect of the optimized cuts of Table~\ref{tabcut}
in Fig.~\ref{bacuts}, where the cluster tranverse mass distribution
for a $\mh=170$ GeV signal and
the summed backgrounds, normalized to 30 $\fbi$, are shown 
before (a) and after the final cuts (b). A clear excess
of events from the Higgs signal can be seen in
Fig.~\ref{bacuts}(b). 
It is important to note that before application of the
final cuts, the dominant backgrounds are $WW$ and the $W$+fake
with other sources accounting for less than 10\% 
of the total. Moreover, for 30 $\fbi$ integrated luminosity, 
the statistical error in the background is less than 2\% 
before application of the final cuts. 
We therefore argue that one should be able 
to normalize the SM background curve ($WW$) with sufficient
precision to unambiguously identify a significant excess 
attributable to Higgs boson signal. It should also be noted that by
selectively loosening the final cuts, it is possible 
to maintain the same $S/\sqrt{B}$ while increasing the accepted 
background by up to factor of 5, and the accepted signal
by a factor of 2.5. This can provide a powerful cross-check
of the predicted background $M_C$ shape and can be used
to demonstrate the stability of any observed excess.

\begin{figure}[thb]
\epsfysize=5.5in
\epsffile[0 100 560 560]{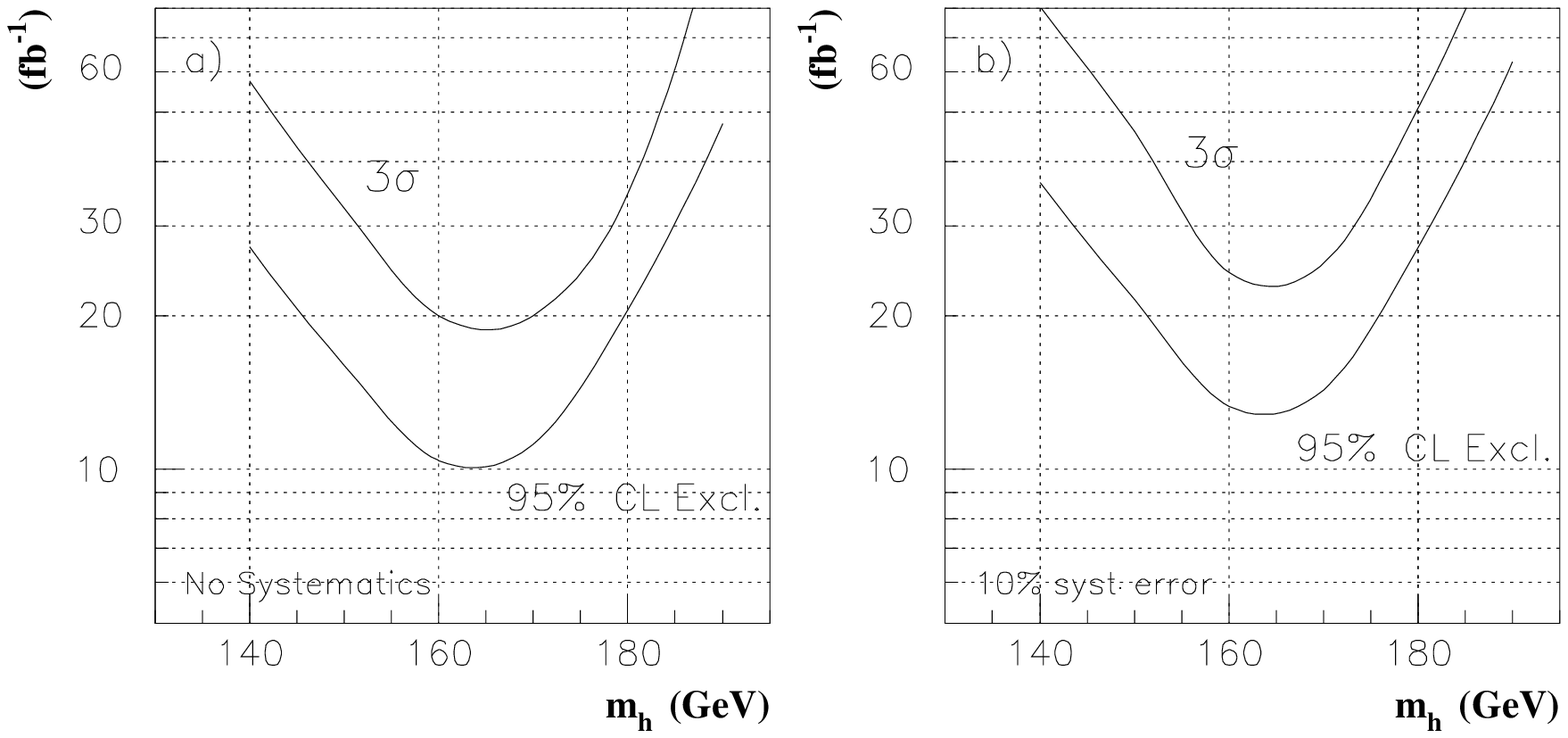}
\vskip -2.0in
\caption[]{The integrated luminosity required to reach $3\sigma$
statistical significance and 95\%
exclusion versus $m_h$ in the $h\to \ww \to \ell\bar \nu
\bar \ell \nu$ channel for (a) statistical effects only; (b) 10\%
systematic error for the signal and SM backgrounds included. 
The contribution from $W\to \tau \to \ell$ decays, associated 
production and gauge boson fusion have also been included.
\label{intL}}
\end{figure}

Our final results for the channel
$h\to \ww\to \ell \bar \nu \bar \ell \nu$ are summarized in
Table~\ref{lnln}. We have included the contributions to $h\to \ww$
from the signal channels in Eqs.~(\ref{whzh}) and (\ref{wwzz}).
Although they are small to begin with, they actually increase 
the accepted signal cross section by $12-18\%$. 
We have also 
included the contribution from 
$W \to \tau \nu \to \ell\nu_\ell \nu$.\footnote{From consideration 
of the $W$+$(j\rightarrow e)$ background,
it should be clear that improving the sensitivity 
by incorporating hadronic tau decays will be a difficult task.
We nonetheless encourage the effort.}
It can be seen that one may achieve a $S/B$ of at least 6\% 
for 140 GeV$<\mh< 190$ GeV and reach 45\% 
for $\mh=170$ GeV. The statistical significance,
$S/\sqrt B$, for 30 fb$^{-1}$ integrated luminosity, 
is 3$\sigma$ or better for $150< \mh < 180$ GeV.
In Fig.~\ref{intL}(a), we present the integrated luminosities 
needed to reach a 3$\sigma$ significance and 95\% 
CL exclusion computed assuming Poisson probabilities for $\mh$.

\begin{table}[thb]
\begin{tabular}{|l|c|c|c|c|c|c|}
 $\mh$ [GeV] & 140 & 150 & 160 & 170 & 180   & 190\\  \hline
 $gg\to h$ [fb] & 2.2 & 2.4 & 1.3 & 0.93 & 0.85 & 0.73\\ 
 associated $VH$ [fb] & 0.26 & 0.31 & 0.13 & 0.09 & 0.06   & 0.06\\ 
 $VV$ fusion [fb] & 0.12 & 0.12 & 0.09 & 0.06 & 0.05 & 0.05\\ 
 signal sum [fb] & 2.6 & 2.8 & 1.5 & 1.1 & 0.96 & 0.83\\ 
\hline
SM bckgrnds [fb]& 39 & 27  & 4.1 & 2.3 & 3.8 & 7.0 \\ 
fake $j\to e$ [fb]& 5.1 & 3.4  & 0.34 & 0.15 & 0.08 & 0.45 \\ 
bckgrnds sum [fb]& 44 & 30  & 4.4 & 2.4 & 3.8 & 7.5 \\ \hline
$S/B$ [\%] & 5.8 & 9.4  & 34 & 45 & 25 & 11 \\ \hline
$S/\sqrt B$ [30 fb$^{-1}$] & 2.1 & 2.8  & 3.9 & 3.8 & 2.7 & 1.7 \\ 
\end{tabular}
\vspace{0.2in}
\caption[]{Summary table for $h\rightarrow \ww \rightarrow \ll$ signal 
for $\mh=140-$190 GeV and various SM backgrounds after the kinematical 
cuts of Eqs.~(\ref{basic})$-$(\ref{jetveto}) and the likelihood cut
Eq.~(\ref{likec}). $W+$fake refers to the background where a jet mimics
an electron with a probability of $P(j\to e)=10^{-4}$.
The backgrounds are independent of $\mh$.
}
\label{lnln}
\end{table}

To assess the effect of inherent systematic uncertainties,
we re-evaluate the corresponding curves
in Fig.~\ref{intL}(b) assuming a 10\% 
systematic error for the signal and SM 
backgrounds.\footnote{For the purposes of computing the effects of 
systematic errors on the sensitivity to a Higgs signal, 
we have scaled the expected background upward by a given percentage
and the expected signal downward by the same percentage simulataneously.} 
The results are somewhat degraded, but they are still encouraging.

\section{Like-sign Di-lepton Plus Jets Signal}

When considering the $h\to \ww$ mode in the associated production
channels of Eq.~(\ref{whzh}), it is natural to consider the trilepton
mode of Eq.~(\ref{unlikesign})  \cite{baerwells}. However, the 
leptonic branching fractions for the $W$ decays limit the signal rate. 
Also, the leading irreducible SM background 
$WZ(\gamma^*) \to 3\ell $ is difficult
to suppress to a sufficient level. 
On the other hand, the $\ww \to \ell\nu jj$ mode gives like-sign
leptons plus two-jets events \cite{scott,baerwells} as 
in Eq.~(\ref{likesign}) with a three times larger rate
than the trilepton mode; while the leading background
is higher order than $WZ(\gamma^*)$.
In this case, the contributing channels include
\begin{eqnarray}
&&Wh \to W\ww \to \ell^\pm\nu \ell^\pm\nu jj,\nonumber\\
&&Wh \to W\zz \to \ell^\pm\nu \ell^\pm\ell^\mp jj,\nonumber\\
&&Zh \to Z\ww \to \ell^\pm\ell^\mp \ell^\pm\nu jj,\nonumber\\
&&Zh \to Z\zz \to \ell^\pm\ell^\mp \ell^\pm\ell^\mp jj.\nonumber
\end{eqnarray}
We identify the final state signal as two isolated 
like-sign charged leptons plus jets. A soft third lepton
may be present. The SM backgrounds are
\begin{eqnarray}
\label{vvv}
p\bar p &\to& WWW,\ WWZ,\ WZZ,\ ZZZ,\ t \bar t W,\ t \bar t Z
\to \ell^\pm \ell^\pm jj\  X,\\
\label{wzjj}
p\bar p &\to& W^\pm Z(\gamma^*)+jj\to  \ell^\pm \ell^\pm jj X,\ \ 
Z Z(\gamma^*)+jj\to \ell^\pm \ell^\pm jj X,\ \  
t \bar t\to  \ell \bar \nu jj b\bar b,\\
p\bar p &\to& Wjj,\ Z(\gamma^*)jj +{\rm fake}.\nonumber
\end{eqnarray}
Although the triple gauge boson production \cite{hs}
in Eq.~(\ref{vvv}) constitutes the irreducible backgrounds,
the $WZjj$, $t\bar t$ through $b$ or $c$ semileptonic decay
and the background from $j\to e$ fakes turn out to be larger.

\begin{figure}[tb]
\epsfysize=5.5in
\epsffile[-50 0 560 560]{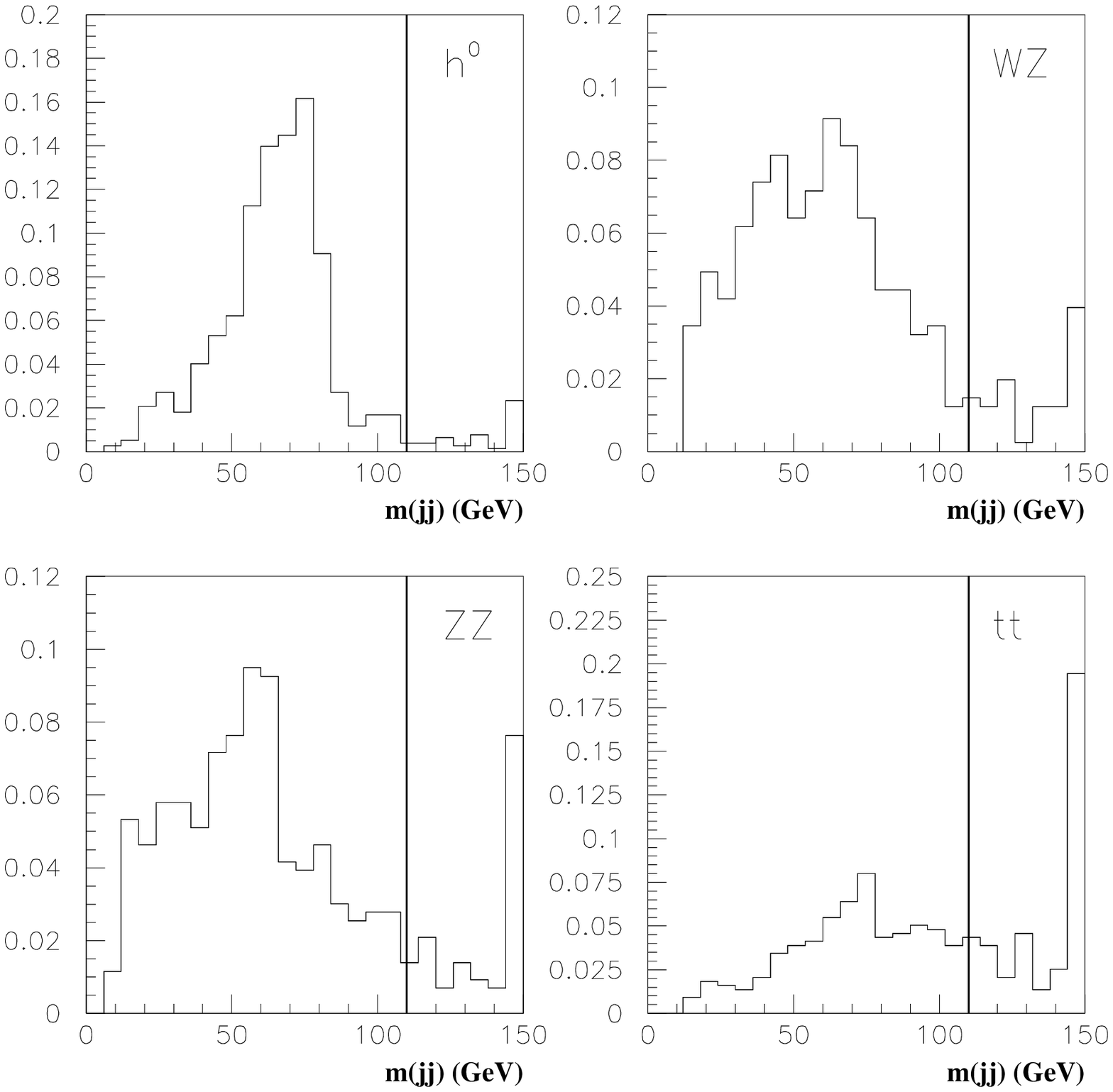}
\caption[]{Normalized di-jet mass distributions 
${1\over \sigma} {d\sigma\over d m(jj)}$ for the signal 
$W^\pm h \to W^\pm \ww \to \break  \ell^\pm\ell^\pm jj$ 
with $\mh=170$ GeV and the backgrounds $WZ$, $ZZ$ and $t\bar t$.
\label{mjj}}
\end{figure}

The basic acceptance cuts required for the leptons are 
\begin{eqnarray}
\nonumber
&&p^{}_T(\ell)  > 10\ {\gev},\ \ |\eta^{}_\ell| < 1.5,\ \ 
m(\ell\ell)>10\ {\gev},\nonumber\\
&&0.3 < \Delta R(\ell j) <6,\ \ \etmiss > 10\ {\gev}.
\label{basic2}
\end{eqnarray}
For a muon, we further demand that the scalar sum of additional
track momenta within 30$^\circ$ be less than 60\%
of the muon momentum.
We require that there are at least two jets with
\begin{equation}
p_T^j>15\ {\gev},\ \  |\eta^{}_j| < 3.
\label{jets}
\end{equation}
To suppress the $WZ$ background, we require 
the leading jet to be within $|\eta^{}_{j_1}| < 1.5$
and to have a charged track multiplicity satisfying
$2\leq N\leq 12$;
while the sub-leading jet to be within $|\eta^{}_{j_2}| < 2.0$.
The $t\bar t$ background typically exhibits greater jet activity;
we therefore veto events having 
\begin{equation}
 p_T^{j_3}>30\ {\gev},
\end{equation}
and events with a fourth jet satisfying Eq.~(\ref{jets}).
To suppress backgrounds associated with heavy flavor jets, 
we veto the event if any of the jets have a $b$-tag. 

In Fig.~\ref{mjj}, we present the di-jet mass distributions
for the signal and backgrounds. Since the di-jets in the signal
are mainly from a $W^*$ decay, $m(jj)$ is close to or lower than
$M_W$. This motivates us to further require
\begin{equation}
m(jj)<110\ {\gev},\ \ \Sigma_j |p_T^j|<150\ {\gev}.
\label{mpt}
\end{equation}

\begin{figure}[tb]
\epsfysize=5.5in
\epsffile[-50 0 560 560]{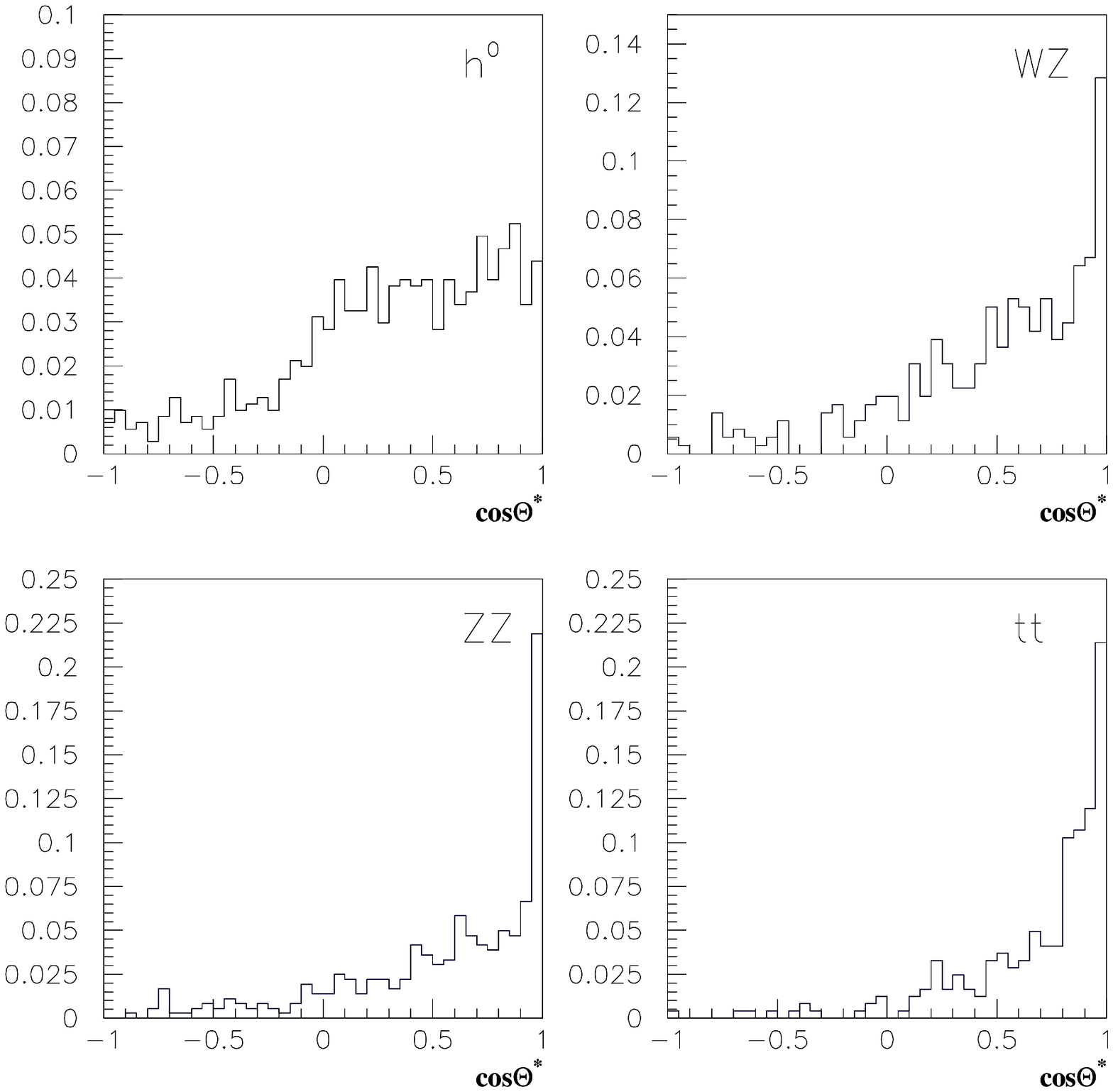}
\caption[]{Normalized distributions 
${1\over \sigma} {d\sigma\over d \cos\theta^*_{\ell_1} }$ 
for the correlation angle defined
above Eq.~(\ref{theta}) for the signal 
$W^\pm h \to W^\pm \ww \to \ell^\pm \ell^\pm jj$ 
with $\mh=170$ GeV and backgrounds $WZ$, $ZZ$ and $t \bar t$. 
\label{thetass}}
\end{figure}

Finally, it is interesting to note that the lepton correlation
angle introduced in Eq.~(\ref{theta}) has strong discriminating
power to separate the signal from backgrounds as shown in 
Fig.~\ref{thetass}. We then impose a final cut
\begin{equation}
\cos\theta^*_{\ell_1} < 0.95.
\end{equation}

\begin{table}[thb]
\begin{tabular}{|l|c|c|c|c|c|c|c|c|c|}
 $\mh$ [GeV] &120 &130 & 140 & 150 & 160 & 170 & 180 & 190 &200\\ \hline
 signal sum [fb] &0.093 &0.20 &0.34 &0.52 &0.45 & 0.38& 0.29 &0.20 &0.16\\ 
\hline
\hline
bckgrnd channels&$WZ$&$ZZ$&$WW$&$t\bar t$&$VVV$ &$t\bar t V$ &
$W/Z\ jj+$fake & Sum  &\\
$\sigma$ [fb]& 0.27&0.06 &0.01&0.15&0.07 &0.02 &0.26 \cite{w3jet}&0.83 &\\ 
\hline
\hline
$S/B$ [\%] & 11 & 24  & 41 & 63 & 54 & 46 & 35 & 24& 19 \\ \hline
$S/\sqrt B$ [30 fb$^{-1}$] &0.56 &1.2 &2.0 &3.1 &2.7 &2.3&1.7&1.3&0.96\\ 
\end{tabular}
\vspace{0.2in}
\caption[]{$V h \to \ell^\pm\ell^\pm jj$ 
signal for $\mh=120-$200 GeV and the SM backgrounds 
after the kinematical cuts of Eqs.~(\ref{basic2})$-$(\ref{mpt}). 
}
\label{lljj}
\end{table}

With these cuts, we present the results for the signal
and backgrounds in Table~\ref{lljj}. We can see that for a
given $\mh$, the $S/B$ is larger than that for the di-lepton
plus $\etmiss$ signature, reaching as high as 63\%. One can consider
further optimization of cuts with $\mh$ dependence.
However, the rather small signal rate for a 30 fb$^{-1}$
luminosity limits the statistical significance.
Also, the systematic uncertainty in the background 
may be worse than the pure leptonic channel.

\section{Discussions and Conclusion}

\begin{figure}[tb]
\epsfysize=5.5in
\epsffile[0 100 567 567]{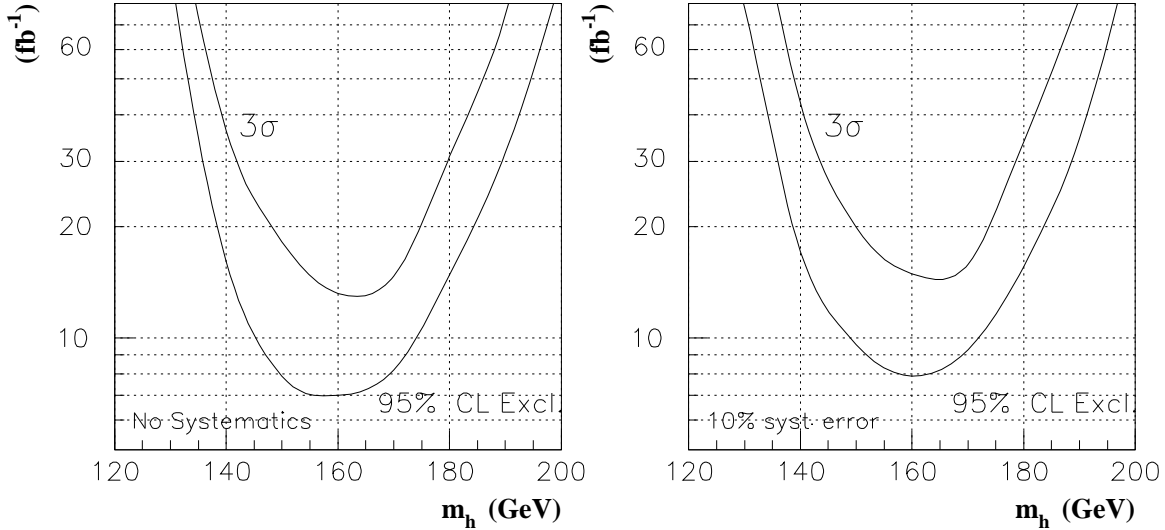}
\vskip -2.0in
\caption[]{The integrated luminosity required to reach $3\sigma$
statistical significance and 95\% 
CL exclusion 
versus $m_h$ in combining both channels 
$h\to \ww \to \ell\bar \nu \bar \ell \nu$ and
$W^\pm h\rightarrow \ww,\zz \to \ell^\pm\ell^\pm jj$
for (a) statistical effects only; (b) 10\%
systematic error included for these two channels.
\label{intLF}}
\end{figure}

We have carried out comprehensive studies for $h\to \ww$
via the two channels
\begin{eqnarray}
\label{purell}
&& p\bar p\to h\to \ww \to \ell\bar \nu \bar \ell \nu,\\
&& p\bar p \to W^\pm h,\ Zh\to W^\pm(Z)\ww \to \ell^\pm\ell^\pm jj.
\label{llajj}
\end{eqnarray}
In combining both channels,
we present our summary figure in Fig.~\ref{intLF},
again for (a) statistical effects only; (b) 10\%
systematic error for the signal and SM backgrounds
included for both channels.
We conclude that with a c.~m.~energy of 2 TeV and an integrated
luminosity of 30 fb$^{-1}$ the Higgs boson signal via
$h\to \ww$ should be observable at a 
3$\sigma$ level or better for the mass range of 
$145\ {\gev} \lsim \mh \lsim 180$ GeV. For 95\% 
CL exclusion, the mass reach is
$135\ {\gev} \lsim \mh \lsim 190$ GeV. 

\begin{figure}[thb]
\epsfysize=5.5in
\epsffile[0 100 560 560]{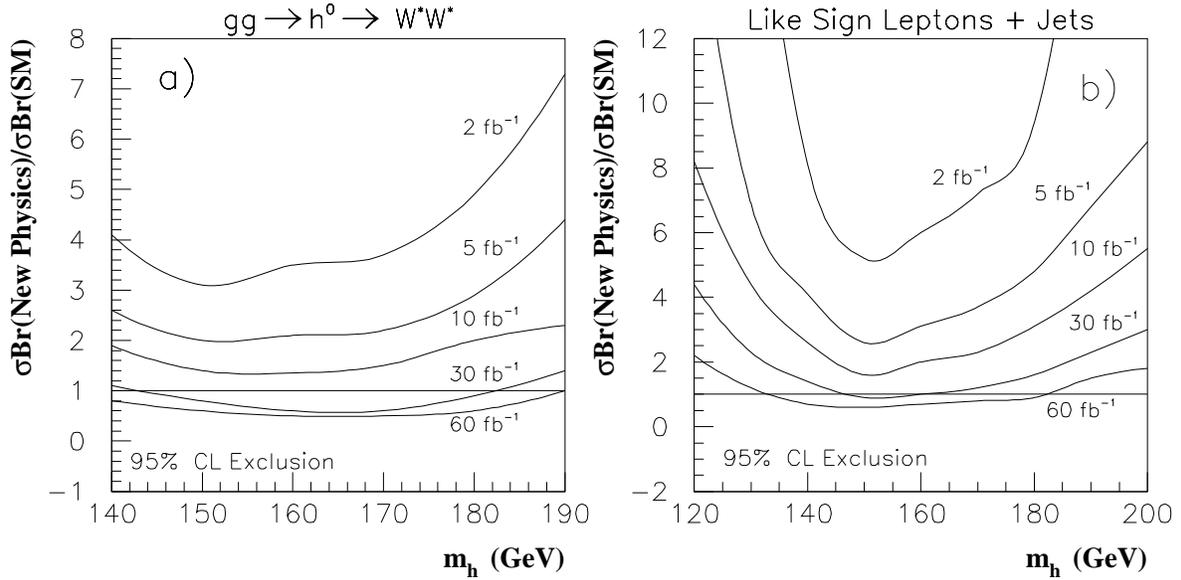}
\vskip -2.0in
\caption[]{95\% 
CL exclusion for the ratio $R$ [Eq.~(\ref{R})]
versus $\mh$ for several values of the integrated luminosity, 
(a) for the channel Eq.~(\ref{purell}), and (b) for 
Eq.~(\ref{llajj}). 
\label{generic}}
\end{figure}

Our results presented here are valid not
only for the SM Higgs boson, but also for
SM-like ones such as the lightest supersymmetric Higgs boson
in the decoupling limit \cite{decouple}.
A Higgs mass bound can be translated into exploring fundamental
parameters for a given theoretical model, as shown in 
Ref.~\cite{baeretal}. 
Furthermore, if there is an enhancement for 
$\Gamma(h\to gg)\times BR(h\to WW,ZZ)$ over 
the SM expectation, or if $BR(h\to b\bar b)$
is suppressed, such as in certain parameter
region in SUSY \cite{CMW}, the signals of Eq.~(\ref{vv})
would be more substantial and more valuable to study.
We can make our study more general in this regard
by considering the quantity $\sigma(h)\times BR(h\to \ww)$ 
as a free parameter. Define a ratio of this parameter 
to the SM expectation for the signal to be
\begin{equation}
R={\sigma(h)\times BR(h\to \ww)_{New Physics}
\over \sigma(h)\times BR(h\to \ww)_{SM}}.
\label{R}
\end{equation}
Measuring $R$ would represent a generic Higgs boson search in
a model-independent way. Figure~\ref{generic} gives the 95\%
CL exclusion for the ratio $R$ versus $\mh$ for several 
values of the integrated luminosity, where $R=1$ corresponds
to the SM expectation. Figure~\ref{generic}(a) is for the
channel Eq.~(\ref{purell}) where we have only included
the gluon-fusion contribution, 
and Fig.~\ref{generic}(b) for 
Eq.~(\ref{llajj}). On the other hand, once a Higgs boson 
signal is established, a careful examination of $R$ would
help confirm the SM or identify possible new physics.

Finally, we would like to point out that further improvement
on our results is still possible by including other channels.
Although there would be even larger SM backgrounds, the channel 
$h \to \ww \to \ell \nu jj$ was found \cite{hz}
to be helpful in improving the Higgs boson coverage. 
Combining with  $h \to Z^*Z^* \to \ell\ell jj$ as shown
in Fig.~\ref{two}, we would expect some possible 
improvement which deserves further study.
The channels $h \to Z^*Z^* \to \ell\bar \ell \nu\bar \nu, 4\ell$
may have smaller SM backgrounds, especially for
the $4\ell$ mode. Unfortunately,
the signal rate would be very low for the anticipated
luminosity at the Tevatron. It is nevertheless prudent
to keep them in mind in searching for the difficult
Higgs boson signal. 

In summary, we have demonstrated the feasibility for an
upgraded Tevatron to significantly extend the Higgs boson
mass coverage. The Fermilab Tevatron with 
luminosity upgrade will have the potential 
to significantly advance our knowledge of Higgs boson physics.

\vskip 0.1in

{\it Acknowledgments}: 
We would like to thank the participants of the Tevatron Run-II
Higgs Working Group, especially M. Carena, J. Conway, H. Haber,
J. Hobbs, J.-M. Qian, D. Rainwater, S. Willenbrock 
and D. Zeppenfeld for discussions.
We would also like to thank T. Junk for making available the
programs used to calculate the sensitivity in the Poissonian limit,
and E. Flattum for developing the SHW/PAW interface.
This work was supported in part by a DOE grant No. 
DE-FG02-95ER40896 and in part by the Wisconsin Alumni 
Research Foundation. A.S.T acknowledges the support from
the University of Michigan by a DOE grant No. DE-FG02-95ER40899.

\end{document}